\documentclass[a4paper,11pt]{article}
\usepackage{amsmath, amssymb}
\usepackage{type1cm}
\usepackage{bm}
\usepackage{comment}
\usepackage[dvipdfmx]{graphicx}
\usepackage{float}
\usepackage{graphicx}
\usepackage{tikz}
\usepackage{amssymb,mathtools}
\usetikzlibrary{arrows.meta,calc}

\newcommand{\ctext}[1]{\raise0.2ex\hbox{\textcircled{\scriptsize{#1}}}}

\title{From Three-Particle Dynamics to the Structural Origin
of the Arrow of Time in Classical and Quantum Mechanics}
\author{Shuhei Kobayashi%
\thanks{Corresponding author: \texttt{xiaolinxiuping1@gmail.com}}}

\date{December 18, 2025} 
\begin{document}
\maketitle
\begin{center}
Department of Physics and Astronomy, Faculty of Science and Technology, Tokyo University of Science, Noda City, Chiba, Japan\\
\texttt{6222041@ed.tus.ac.jp} \quad (long-term: \texttt{xiaolinxiuping1@gmail.com})
\end{center}
\begin{abstract}

This paper presents a unified formulation of the origin of the arrow of time in classical and quantum mechanics. We begin with a mechanical analysis of a one-dimensional three-particle system, which provides a concrete example in which macroscopic irreversibility emerges despite microscopically reversible dynamics. By abstracting this mechanism, we identify coarse-graining as the essential ingredient responsible for macroscopic time asymmetry.

We then formulate a general structural criterion for the thermodynamic arrow of time. We show that when microscopic time evolution forms a group while the induced macroscopic evolution forms only a semigroup, macroscopic time-reversal symmetry is necessarily broken. We prove that this semigroup structure arises if and only if the coarse-graining map from microscopic to macroscopic states is non-injective.

This result holds independently of whether the underlying system is classical or quantum. In the quantum case, using density matrices, antiunitary time reversal, and CPTP coarse-graining maps, we show that macroscopic irreversibility follows inevitably from information loss, without requiring any asymmetry in the microscopic laws.

Our results demonstrate that the thermodynamic arrow of time has a universal structural origin: the loss of microscopic information inherent in coarse-graining.

\end{abstract}
\section{Mechanical Analysis}
The velocities of each particle in the three-particle system after any number of collision is known.$^{[1]}$ As widely known, the temperature of a system is propotion to the average of molecules' kinetic energy. Thus, we represented particle A and C as finite thermal baths and B as thermal wall and expressed A being hotter than C at initial condition with $|v_{A0}|>|v_{C0}|$. Plus, from previous paper, the expression of velocities of each particle have a term $\overline{v_{AB}}=\frac{v_{A0}+v_{B0}}{2}$ and this makes equations more complicated. Therefore, for simplicity, we set
\begin{align}
v_{B0}=-v_{A0}=u>0
\end{align}
so that the collisions between A and B can occur and the average of them is 0. In addition, we set the initial velocity of C as
\begin{align}
v_{C0}=-v<0
\end{align}
so that the temperature of C increases as the collisions happen between B and C.By substituting these into previous paper's formula, we get
\begin{align}
\begin{bmatrix}
v_{A2k}\\
v_{B2k}\\
v_{C2k}
\end{bmatrix}
=
\begin{bmatrix}
\left(\nu^2 v + c u\right)\frac{\sin{k\theta}}{d} - u\cos{k\theta}\\
-\left[\left(\nu^2 v + c u\right)\frac{\sin{k\theta}}{d} - u\cos{k\theta}\right]\\
-(u + c v)\frac{\sin{k\theta}}{d} - v\cos{k\theta}
\end{bmatrix}
\end{align}

\begin{align}
&\begin{bmatrix}
v_{A2k+1}\\
v_{B2k+1}\\
v_{C2k+1}
\end{bmatrix}
=\frac{1}{M_{AB}}
\begin{bmatrix}
\left(m_A-3m_B\right)\left[\left(\nu^2v+cu\right)\frac{\sin{k\theta}}{d}-u\cos{k\theta}\right]\\
\left(m_B-3m_A\right)\left[\left(\nu^2v+cu\right)\frac{\sin{k\theta}}{d}-u\cos{k\theta}\right]\\
-M_{AB}\left[\left(u+cv\right)\frac{\sin{k\theta}}{d}-v\cos{k\theta}\right]
\end{bmatrix}
\end{align}
Now, we define below;
\begin{align}
\varepsilon_A=\frac{m_B}{m_A}\\
\varepsilon_C=\frac{m_B}{m_C}
\end{align}
In this case, the mass of B is smaller enough than other particles, which leads that the amount of these will be $<<$1. At the same time, $c,d,\nu^2$ will be
\begin{align}
c\approx 2\varepsilon_A\\
d\approx\frac{\varepsilon_A}{\sqrt{\varepsilon_C}}\\
\nu^2\approx1-(\varepsilon_A+\varepsilon_C)
\end{align}
for first order approximation. Because both $\varepsilon_A$ and $\varepsilon_C$ will be the same scale, $d$ will be closer to 0 when these amounts are small enough. Plus, the total number of collisions depends on $\theta$ therefore this also tends to 0 increasing the number of collisions sufficiently for a thermodynamic approximation. Thus,
\begin{align}
\frac{\sin{k\theta}}{d}\rightarrow k\\
\cos{k\theta}\rightarrow 1
\end{align}
Based on these, we get
\begin{align}
\begin{bmatrix}
v_{A2k}\\
v_{B2k}\\
v_{C2k}
\end{bmatrix}
\approx
\begin{bmatrix}
k\{[1-(\varepsilon_A+\varepsilon_C)]v+2\varepsilon_Au\}-u\\
-k\{[1-(\varepsilon_A+\varepsilon_C)]v+2\varepsilon_Au\}+u\\
-k(2\varepsilon_Av+u)-v
\end{bmatrix}
\end{align}
Now we calculate the ratios and differences of energy. First, the energy ratio of A and B boils down to the mass ratio because the square of velocities matches. Thus, the energy ratio is $\varepsilon_A$ itself and this is $<<$1 so that the energy of A is rarely transfered to B Plus, In this approximation, B is much lighter than A and C thus the relative velocities of B from A and C sill be greater. Therefore, the time taken for each cycle is approximately constant. This result shows that the mass of B is small enough and as a coclusion B can be a good channel. This is connected as thermal wall having small enough heat capacity, which let the heat from hotter bath to cooler bath be transfered almost directly based on the thermodynamic analogy. Thus, we let the time B going back and forth between A and C be a constant $\Delta t$. Now, from the law of conservation of momentum, the change of velocities through the collision of A and B under the approximation of masses are calculated as 
\begin{align}
v_A'=\frac{m_A-m_B}{M_{AB}}v_A+\frac{2m_B}{M_{AB}}v_B\approx v_A+2\varepsilon_A(v_B-v_A)\\
v_B=\frac{2m_A}{M_{AB}}v_A+\frac{m_B-m_A}{M_{AB}}v_B\approx 2v_A-v_B+2\varepsilon_A(v_A-v_B)
\end{align}
with velocities of each particle before and after the collision $v_A,v_B$ and $v_A',v_B'$.
Therefore, the change of energy of A $\Delta E_A$ is
\begin{align}
\Delta E_A=\frac{1}{2}m_A\left(v'^2-v^2\right)\notag\\
\approx 2m_Bv_A(v_B-v_A)
\end{align}
Similarly in the collision of B and C, the change of energy of C is, under the first order approximation, 
\begin{align}
\Delta E_C\approx 2m_Bv_C(2v_A-v_B-v_C)
\end{align}
From this, the difference between the change of A and C after a cycle is expressed
\begin{align}
\Delta E_A-\Delta E_C\approx 2m_B(v_A-v_C)(v_B-v_A-v_C)
\end{align}
with the velocities before the cycle starts $v_A,v_B,v_C$. Therefore, the differences in the change of enregy of A and C at $k$-th cycle is
\begin{align}
\Delta E_A^{(k)}-\Delta E_C^{(k)}\approx 2m_B(v_{A2k}-v_{C2k})(v_{B2k}-v_{A2k}-v_{C2k})\notag
\end{align}
Here, we define
\begin{align}
\Delta_k\equiv E_A^{(k)}-E_C^{(k)}\\
\Lambda_k=\frac{\Delta E_A^{(k)}-\Delta E_C^{(k)}}{\Delta_k}
\end{align}
and this leads this approximation;
\begin{align}
\Delta_{k+1}-\Delta_k\approx-\Lambda_k\Delta_k
\end{align}
Based on this, we make 
\begin{align}
\frac{\Delta_{k+1}-\Delta_k}{\Delta t}\approx-\frac{\Lambda_k}{\Delta t}
\end{align}
being continuous and lead
\begin{align}
\frac{\mathrm{d}\Delta}{\mathrm{d}t}\approx-\Gamma_k\Delta
\end{align}
where
\begin{align}
\Gamma_k\equiv\frac{\Lambda_k}{\Delta t}
\end{align}
As $\Gamma_k\approx\Gamma$, we assume $\Gamma_k$ as a constant amount and solve the differential equation to get this;
\begin{align}
\Delta(t)=\Delta(0)e^{-\Gamma t}
\end{align}

which means the differences of energy of two particles drops to exponential relaxation. This indicates the loss of the time reversal symmetry and implication of the flow of time going along only one direction with only the assumption of mass approximation. This apparent irreversibility arises when only the energy (not the direction of velocity) is observed. With the conversion to reverse the direction of time $t\to -t$, the velocity is also conversed similarly $v\to -v$, however, the square of it, which loses the information of sign, is unchanged between before and after the conversion. Therefore, in classical mechanics, which we directly observe the velocity, the time reversal symmetry is kept but in thermodynamics, which the energy is the target of observation, it looks lost. 
\section{Thermodynamic Analysis}

We next assume this system as finite baths connecting. In this situation, in heat equation, the heat capacity of B can be almost neglected and the energy transfered between A and C is rarely absorbed into B. That is why this system is considered as connection of two finite bath through thermal wall. the heat equations are below;
\begin{align}
C_A\dot{T}_A(t)=G(T_A(t)-T_C(t))\\
C_C\dot{T}_C(t)=-G(T_A(t)-T_C(t))
\end{align}
where $G$ is heat conductance and $C$ is heat capacity. The solution of these are
\begin{align}
T_A(t)=T_{eq}-\frac{C_C}{C_A+C_C}\Delta T(0)\exp{\left[-G\left(\frac{1}{C_A}+\frac{1}{C_C}\right)t\right]}\notag\\
T_C(t)=T_{eq}+\frac{C_A}{C_A+C_C}\Delta T(0)\exp{\left[-G\left(\frac{1}{C_A}+\frac{1}{C_C}\right)t\right]}
\end{align}
where $T_{eq}$ is the temperature at the thermal equilibrium state and $\Delta T(0)$ is the initial temperature difference. Therefore, calculating the difference of these, we get
\begin{align}
T_A(t)-T_C(t)=-\Delta T(0)\exp{\left[-G\left(\frac{1}{C_A}+\frac{1}{C_C}\right)t\right]}
\end{align}
which has the same structure asexponential relaxation appeared in mechanical analysis.\\
Here, the temperatures of each particle is, because the degree of freedom is 1, connected with $E_i$ for particle $i$($i$=A,B,C)  based on law of equipartition of energy;
\begin{align}
E_i=\frac{k_BT_i}{2}\\
\therefore T_i=\frac{2E_i}{k_B}=\frac{m_iv_i^2}{k_B}
\end{align}
This implies that we can calculate the instantaneous temperature with only the mass and velocity without depending on the average of physical quantities in case of system with few particles.
\section{Statistical Mechanical Analysis}
Now, we analyze this system from the perspective of statistical mechanics. According to Gibbs-Duhem's equation,
\begin{align}
\frac{\partial S_i}{\partial E_i}=\frac{1}{T_i}\\
\therefore S_i=\int\frac{k_B}{2E_i}\mathrm{d}E_i\\
=\frac{k_B}{2}\ln{|E_i|}+C
\end{align}
therefore, the ratio of entropy generation is 
\begin{align}
\dot{S}_i=\frac{k_B}{|v_i|}
\end{align}
B has much great $|v_i|$ thus the total ratio of entropy generation of A and C is 
\begin{align}
\dot{S}_A+\dot{S}_C=k_B\left(\frac{1}{|v_A|}+\frac{1}{|v_C|}\right)>0
\end{align}
Therefore, the law of entropy increase holds even within a purely mechanical system.
\section{Discussion on This system}

In this study, we derived an effective thermal description from a
completely mechanical three-particle system. The comparison between
mechanical and thermal systems is summarized as follows:

\begin{table}[h]
\centering
\begin{tabular}{ll}
\hline
Mechanical quantity & Thermal analogue \\
\hline
Energies $E_A$, $E_C$ & Internal energies of heat baths A and C \\
Particle $B$ & Thermal wall (mediator of energy) \\
One collision cycle & Infinitesimal time step $dt$ \\
Energy difference $\Delta = E_A - E_C$ & Temperature difference $\Delta T = T_A - T_C$ \\
Coefficient $\Gamma$ & Effective conductance $G(1/C_A + 1/C_C)$ \\
Loss of sign $(v \rightarrow v^2)$ & Coarse-graining / irreversibility \\
\hline
\end{tabular}
\end{table}

When the mediator particle is much lighter than the other two
($m_B \ll m_A, m_C$), the energy difference between A and C decreases
monotonically. The obtained differential equation
\begin{equation}
\frac{d\Delta}{dt} = -\Gamma \Delta
\end{equation}
is equivalent to the thermal relaxation equation of two finite heat
baths in contact,
\begin{equation}
\frac{d(T_A - T_C)}{dt}
= -G\left(\frac{1}{C_A} + \frac{1}{C_C}\right)(T_A - T_C),
\end{equation}
whose solution is the exponential form
\begin{equation}
\Delta T(t) = \Delta T(0)e^{-\Gamma t}.
\end{equation}
This result shows that the macroscopic thermal relaxation can be described by
the microscopic dynamics of elastic collisions.

Although the equations of motion are time-reversal symmetric,
the transformation from $(v_A,v_B,v_C)$ to $(E_A,E_B,E_C)$ removes
the sign of each velocity.
As a result, the macroscopic description based on energy
becomes irreversible. In this sense, the breaking of time-reversal
symmetry arises not from the dynamics themselves but from the loss of
microscopic information caused by coarse-graining.

From the definition of entropy
\begin{equation}
S_i = \frac{k_B}{2}\ln E_i + \text{const},
\end{equation}
and the energy conservation $\dot E_A + \dot E_C \simeq 0$, the total
entropy production rate is
\begin{equation}
\dot{S}_A+\dot{S}_C=k_B\left(\frac{1}{|v_A|}+\frac{1}{|v_C|}\right)>0
\end{equation}

Therefore, the entropy of the system always increases, even though
the underlying dynamics are reversible.

In conclusion, the present model demonstrates that the arrow of time
can appear in a deterministic mechanical system when information about
velocity direction is lost. This provides a simple and quantitative
example that connects classical mechanics with thermodynamics through
coarse-graining. Further work may extend this model to multi-particle
or quantum systems to explore collective and microscopic origins of
irreversibility.This work thus bridges microscopic reversibility and macroscopic irreversibility within a single deterministic framework.
\section{General Discussion in Classical Mechanics}
We shall now discuss the general origin of arrow of time. Let the degree of freedom of the system be $f$ and generalized coordinate be given for $1\leq i \leq f$ and generalized momentum $p_i(t)$ is shown as below
\begin{align}
p_i(t)=\frac{\partial L}{\partial \dot{q_i}(t)}
\end{align}
where $L$ is lagrangian of the system. Considering a plane $P$ with axes of these generalized coordinate and momentum on phase space, the operation of reversing time corresponds to reflecting a point $(q_i(t),p_i(t))$ on $P$ in $p_i(t)$ axis, thus we can define micro-time-reversal operater $T$ as
\begin{align}
T\left(\begin{matrix}
q_i(t)\\
p_i(t)
\end{matrix}\right)
=
\left(\begin{matrix}
1&0\\
0 &-1
\end{matrix}\right)
\left(\begin{matrix}
q_i(t)\\
p_i(t)
\end{matrix}\right)
=
\left(\begin{matrix}
q_i(t)\\
-p_i(t)
\end{matrix}\right)
\\
\therefore T=\left(\begin{matrix}
1&0\\
0 &-1
\end{matrix}\right)
\end{align}

Plus, we define a set $X$ whose elements are points $x$ in phase space, and
consider a non-injective map $\varphi:X\to Y$.  
Any quantity that takes values in $Y$ is referred to as a macro-variable.
For example, when the Hamiltonian represents the total energy of the system,
the signs of each component of $x$ are ignored, so $f$ distinct points in phase
space can correspond to the same value of the Hamiltonian.
Thus, in this case, the Hamiltonian is an element of $Y$.

Let $y(t)=\varphi(x(t))$ be a macro-variable, and let $\bar{T}$ denote
the time-reversal operation acting on macroscopic variables.
We shall show that
\begin{align}
\bar{T}y(t)\neq\varphi(Tx)
\end{align}
That is
\begin{align}
\bar{T}\phi\neq\varphi T
\end{align}
\section{Theorem}
From the definition, since $\varphi$ is non-injective, one has
\begin{align}
\exists x_1\neq x_2:\varphi(x_1)=\varphi(x_2)
\end{align}
On the other hand, the microscopic time-reversal operation $T$ is clearly injective, so
\begin{align}
x_1\neq x_2\Rightarrow Tx_1\neq Tx_2
\end{align}
Applying $\bar{T}$ to both sides of Eq.~(46), we obtain
\begin{align}
\bar{T}\varphi(x_1)=\bar{T}\varphi(x_2)
\end{align}
However, due to the non-injectivity of $\varphi$, there is in general no guarantee
that the images of $Tx_1$ and $Tx_2$ under $\varphi$ coincide:
\begin{align}
\varphi(Tx_1)=\varphi(Tx_2)
\end{align}
does not necessarily hold.  
Thus, for some pair $(x_1,x_2)$,
\[
\bar{T}\varphi(x_1)) = \bar{T}(\varphi(x_2))
\quad\neq\quad
\varphi(Tx_1),\;\varphi(Tx_2)
\]
This means that, at some points,
\[
\bar{T}\varphi\neq\varphi T
\]
and therefore the two operations cannot be regarded as commuting in general.
Thus we have shown
\[
\bar{T}(y(t)) \neq \varphi(Tx(t))
\]
To clarify the properties of macroscopic variables, note first that sums,
products, and derivatives of macroscopic variables can all be written as
\begin{align}
\varphi'=F\circ\varphi
\end{align}
and therefore are themselves macroscopic variables.
Using this fact, we show that several standard thermodynamic quantities satisfy
the above definition, and that many other macroscopic variables exist.

First, consider temperature.  
From the kinetic theory of gases, the temperature is proportional to the total
kinetic energy of the constituent molecules.
Because kinetic energy depends on the square of the (translational) momenta,
it is an even function of momentum, and therefore a macroscopic variable.

Moreover, the total kinetic energy appearing in this argument is the internal
energy of an ideal gas, and for real gases only potential energy due to
intermolecular forces is added, so internal energy is also a macroscopic
variable by the same reasoning.

Since temperature is defined as the derivative of internal energy with respect
to entropy, entropy must itself be a macroscopic variable.

In addition, volume is determined solely by the difference between the maximum
and minimum positions of the particles.  
Thus it depends only on relative positions and discards detailed coordinate
information, making it a macroscopic variable.

From the thermodynamic identity
\begin{align}
\left(\frac{\partial U}{\partial V}\right)_S=-p
\end{align}
pressure $p$ is also a macroscopic variable.  
The chemical potential $\mu$ and particle number $N$ are likewise manifestly
macroscopic, and therefore any thermodynamic potential obtained from the
internal energy
\begin{align}
U=TS-pV+\mu N
\end{align}
via Legendre transformation is also a macroscopic variable.
\section{Discussion: Physical Interpretation}

In this section, we discuss how the noncommutativity between the
coarse-graining map and the time-reversal operation,
\[
\bar{T}\varphi\neq\varphi T
\]
is consistent with microscopic mechanics and with the thermodynamic arrow of
time.

\subsection{Microscopic reversibility and macroscopic irreversibility}

In classical Hamiltonian mechanics, the time-reversal transformation
$T:(q,p)\mapsto(q,-p)$ is invertible, and the equations of motion are invariant
under this transformation.  
Thus, if $x(t)$ is a microscopic trajectory, the reversed trajectory $x(-t)$ is
also a valid solution, and microscopic dynamics is strictly time-reversal
symmetric.

In contrast, macroscopic variables $y(t)=\varphi(x(t))$ are defined through the
non-injective coarse-graining map $\varphi:X\to Y$, which discards many
microscopic degrees of freedom such as signs, ordering, and local fluctuations.
Therefore, the time-reversal operation and the coarse-graining map do not
commute, and the time evolution of macroscopic variables acquires
\emph{irreversibility} not present at the microscopic level.
Crucially, this irreversibility does not contradict microscopic time-reversal
symmetry.  
Rather, macroscopic irreversibility arises \emph{because} coarse-graining is
imposed on an otherwise reversible microscopic dynamics, leading to a coherent
and consistent picture.

\subsection{Relation to the thermodynamic arrow of time}

The thermodynamic arrow of time is usually expressed through the second law,
$\Delta S\ge 0$.  
However, how this irreversible behavior emerges from reversible microscopic
laws has long been debated, and conventional explanations often rely on specific
assumptions.

The noncommutativity
\[
\bar{T}\varphi \neq \varphi T
\]
derived in this work shows, without invoking any special assumptions, that the
origin of thermodynamic irreversibility lies in coarse-graining itself.

Standard thermodynamic quantities; temperature, internal energy, entropy,
volume, etc. are all coarse-grained variables that heavily average and reduce
microscopic information.  
Thus, under time reversal, the corresponding microscopic state cannot be
reconstructed, and generically
\[
\bar{T}(y(t)) \neq y(-t).
\]
Hence the thermodynamic arrow of time emerges naturally.
In particular, entropy is an extreme coarse-graining,
\[
S = k_{\mathrm B}\ln\Omega,
\]
which compresses an enormous number of microscopic states into a single number,
fully consistent with the framework presented here.
\subsection{Universality of macroscopic irreversibility}

A key feature of this theorem is its universality.  
It applies to any Hamiltonian system where a phase space and a
time-reversal operation can be defined, independent of the specific form of the
potential, the number of particles, or the presence of chaos.

Thus the condition for irreversibility is determined not by microscopic
dynamics but by the introduction of a non-injective coarse-graining map.
This shows that macroscopic irreversibility is not a consequence of any
particular physical mechanism, but rather an unavoidable mathematical feature
arising at a more abstract level.
\section{Quantum Extension}
We now extend the general noncommutativity theorem to quantum mechanics using density matrices and CPTP maps. An expression like a point on phase space in classical mechanics corresponds to density matrix $\rho$ in quantum mechanics. Also, as quantum coarse-graining corresponding to classical coarse-graining $\varphi$, we introduce non-injective CPTP mapping such that 
\[
  \mathcal{C} : \rho \mapsto \sigma
\]
which can be also written as $\sigma(t)=\mathcal{C}(\rho ( t ))$.

In classical mechanics, time reverse is given as flipping the signature of momentum $(q,p)\mapsto(q,-p)$ and this is a reversible transfer on the phase space. In quantum mechanics, complex phase on wave function corresponds to the signature of momentum hence the time reversal is expressed with antiunitary operator with complex conjugate $\Theta$. At this time, time reversing density matrix is defined as
\[
  \mathcal{T}(\rho) := \Theta\, \rho^{*}\, \Theta^{-1}
\]
and is an injective mapping similarly to $T$ in classical mechanics in terms of being reversible.
In this study, we use $\mathcal{T}$ as the quantum analogue of $T$ which corresponds to this injectivity.\\
By definition, $\mathcal{C}$ is an non-injective mapping thus
\begin{align}
\exists \rho_1\neq \rho_2:\mathcal{C}(\rho_1)=\mathcal{C}(\rho_2)
\end{align}

Here, $\mathcal{T}$ is a map whose domain and codomain coincide. On the other hand, $\mathcal{C}$ is a mapping from the microscopic space to the macroscopic space so we have to assume the well-defined time-reverse mapping on the macro-space. Let this be $\bar{\mathcal{T}}:\sigma\rightarrow\sigma$ which is injective and we prove that this cannot commute.
It is obvious that micro-time-reversal operation is injective
\begin{align}
\rho_1\neq \rho_2\Rightarrow \mathcal{T}(\rho_1)\neq \mathcal{T}(\rho_2)
\end{align}
By applying $\bar{\mathcal{T}}$ to both side of equation (53), we obtain
\begin{align}
\bar{\mathcal{T}}(\mathcal{C}(\rho_1))=\bar{\mathcal{T}}(\mathcal{C}(\rho_2))
\end{align}
Since 
$\mathcal{C}$ is non-injective, there is no guarantee that applying the reversible map 
$\mathcal{T}$ beforehand will preserve the equality of the coarse-grained states obtained after 
$\mathcal{C}$ acts.
Coarse-graining in quantum mechanics, implemented by CPTP maps, generally involves discarding environmental degrees of freedom or incorporating decoherence, and thus many distinct microscopic states are typically mapped to the same macroscopic state.
Thus we cannot expect 
\[
  \mathcal{C}(\mathcal{T}(\rho_1))
  = \mathcal{C}(\mathcal{T}(\rho_2))
\]
to hold; in general it is false. Therefore, for some pair of states
\(\rho_1, \rho_2\), one has 
\[
\bar{\mathcal{T}}(\mathcal{C}(\rho_1) )= \bar{\mathcal{T}}(\mathcal{C}(\rho_2))
\quad\neq\quad
\mathcal{C}(\mathcal{T}(\rho_1)),\;\mathcal{C}(\mathcal{T}(\rho_2))
\]

This means that for some states,
\[
\bar{\mathcal{T}}\circ\mathcal{C}\neq\mathcal{C}\circ\mathcal{T}
\]
thus in general we cannot say both can commute.
Because of above, we can say
\[
\bar{\mathcal{T}}(\sigma(t)) \neq \mathcal{C}(\mathcal{T}(\rho(t)))
\]
and it has been proven that coarse mapping and time-reverse operation do not commute in general.\\
From this fact, it becomes clear that both classical and quantum mechanics have the same mathematical origin for the arrow of time. In other words, as stated in discussion of classical mechanics, the origin of arrow of time is based on not microscopic laws or specific mechanism but coarse mapping to generate macro-variable. And this does not contradict time-reversal symmetry in quantum mechanics and rather this phenomenon arises inevitably.

\section{Proof of Necessity and Sufficiency}
In this section, we clarify the role of coarse-graining maps in the generation of the thermodynamic arrow of time and establish its necessary and sufficient condition. To this end, we first redefine the thermodynamic arrow of time. We consider autonomous macroscopic dynamics induced from time-reversal symmetric microscopic laws. We define the thermodynamic arrow of time as the situation in which the microscopic time evolution $\Phi_t$ forms a group, while the induced macroscopic evolution $\Psi_t$ forms only a semigroup.

More precisely, let $f:X\rightarrow Y$ be a map from the set $X$ of microscopic states to the set $Y$ of macroscopic states. Let $\Phi_t$ and $\Psi_t$ denote the time-evolution maps on $X$ and $Y$, respectively. If
\begin{align}
f\circ\Phi_t=\Psi_t\circ f
\end{align}
holds, we identify the semigroup structure of $\Psi_t$ as the thermodynamic arrow of time.

By adopting this definition, we do not discuss irreversibility arising from observational or computational approximations, nor irreversibility associated with extremely low probabilities due to asymmetries in phase-space volume before and after an event. Instead, we assume the existence of an entity, such as Laplace’s demon, that has complete knowledge of all microscopic states and can compute their evolution exactly, and we discuss the irreversibility that such an entity can recognize.

We first consider the case in which the map $f$ is injective. Writing elements of $X$ and $Y$ as $x$ and $y$, respectively, we have $y=f(x)$ and a uniquely defined inverse map $x=f^{-1}(y)$. In this case, the inverse macroscopic evolution
\begin{align}
\Psi_{-t}=f\circ\Phi_{-t}\circ f^{-1}
\end{align}
is well defined. Therefore, $\Psi_t$ forms a group rather than a semigroup, and no thermodynamic arrow of time is generated. By the contrapositive of this result, the existence of a thermodynamic arrow of time requires that the map $f$ is not injective. That is, coarse-graining (a many-to-one map) is a necessary condition for the generation of the arrow of time.

Next, we consider the case in which $f$ is not injective. In this case, distinct microscopic states are mapped to the same macroscopic state, and $f$ is necessarily a many-to-one map. Consequently, an inverse map $f^{-1}$ cannot be defined, and $\Psi_{-t}$ does not exist as the inverse of $\Psi_t$. As a result, the family $\{\Psi_t\}$ forms only a semigroup, and the thermodynamic arrow of time is generated by coarse-graining.

From the above discussion, we conclude that coarse-graining provides a necessary and sufficient condition for the thermodynamic arrow of time. That is, the macroscopic evolution $\Psi_t$ forms a semigroup if and only if the map $f$ is not injective. Therefore, under the assumptions of this section, coarse-graining is the unique structural factor responsible for the thermodynamic arrow of time. This conclusion relies solely on the assumption that the microscopic dynamics forms a group and does not depend on whether the system is classical or quantum. Consequently, even if new physical laws become relevant at previously unexplored scales, the generation of the arrow of time follows from the present argument as long as the microscopic dynamics retains a group structure.

\section{Conclusion}

In this work, we have investigated the origin of the arrow of time from both concrete and abstract perspectives. Starting from a fully deterministic and time-reversal symmetric three-particle system, we demonstrated how macroscopic irreversibility emerges when microscopic information is discarded through coarse-graining.

We then formulated a general structural criterion for the thermodynamic arrow of time. By characterizing microscopic time evolution as a group and macroscopic time evolution as a semigroup, we showed that macroscopic irreversibility arises if and only if the coarse-graining map from microscopic to macroscopic states is non-injective. This establishes coarse-graining as a necessary and sufficient condition for the thermodynamic arrow of time within an autonomous macroscopic description.

Importantly, this result is independent of the specific form of the microscopic dynamics and applies equally to classical and quantum systems. In the quantum case, we demonstrated that the same mathematical structure appears through CPTP coarse-graining maps and antiunitary time reversal, confirming the universality of the mechanism.

These results indicate that the arrow of time does not originate from any particular physical interaction or dynamical asymmetry, but from a universal structural feature: the loss of microscopic information inherent in coarse-graining. As long as microscopic dynamics retains a group structure, macroscopic irreversibility follows inevitably once non-injective coarse-graining is introduced, even in regimes governed by new physical laws.

\end{document}